\documentclass[preprint,aip,jcp,citeautoscript]{revtex4-2}

\usepackage[utf8]{inputenc}
\usepackage{textcomp}
\usepackage[hidelinks]{hyperref}
\usepackage{graphicx}
\usepackage[version=4]{mhchem}
\usepackage{lipsum}
\usepackage{siunitx}
\DeclareSIUnit\au{a.u.}
\usepackage{multirow}
\usepackage{array}
\usepackage{cleveref}
\newcolumntype{?}{!{\vrule width 1pt}}
\newcolumntype{C}[1]{>{\centering\arraybackslash}p{#1}}
\DeclareSIUnit\angstrom{\text{\AA}}

\usepackage{comment}
\usepackage{import}
\allowdisplaybreaks[1]
\usepackage{mathtools,tikz,caption}
\usepackage{ctable}
\usepackage{amssymb}
\usepackage{subcaption} 
\DeclareRobustCommand\sampleline[1]{%
  \tikz\draw[#1] (0,0) (0,\the\dimexpr\fontdimen22\textfont2\relax)
  -- (1.5em,\the\dimexpr\fontdimen22\textfont2\relax);%
}
\colorlet{review}{black}

\begin{document}

\author{Ishna Satyarth}
\affiliation
{Department of Computer Science, Southern Methodist University, Dallas, TX 75275, USA}

\author{Eric C. Larson}
\affiliation
{Department of Computer Science, Southern Methodist University, Dallas, TX 75275, USA}
\author{Devin A. Matthews}
\affiliation
{Department of Chemistry, Southern Methodist University, Dallas, TX 75275, USA}
\email{\{isatyarth,eclarson,damatthews\}@mail.smu.edu}

\title{Tensor Hypercontraction Error Correction Using Regression}

\begin{abstract}
Wavefunction-based quantum methods are some of the most accurate tools for predicting and analyzing the electronic structure of molecules, in particular for accounting for dynamical electron correlation. However, most methods of including dynamical correlation beyond the simple second-order Møller-Plesset perturbation theory (MP2) level are too computationally expensive to apply to large molecules. Approximations which reduce scaling with system size are a potential remedy, such as the tensor hyper-contraction (THC) technique of Hohenstein et al., but also result in additional sources of error.
In this work, we correct errors in THC-approximated methods using machine learning. Specifically, we apply THC to third-order Møller-Plesset theory (MP3) as a simplified model for coupled cluster with single and double excitations (CCSD), and train several regression models on observed THC errors from the Main Group Chemistry Database (MGCDB84). We compare performance of multiple linear regression models and non-linear Kernel Ridge regression models. We also investigate correlation procedures using absolute and relative corrections and evaluate the corrections for both molecule and reaction energies.
We discuss the potential for using regression techniques to correct THC-MP3 errors by comparing it to the “canonical” MP3 reference values and find the optimum technique based on accuracy. We find that non-linear regression models reduced root mean squared errors between THC- and canonical MP3 by a factor of 6-9$\times$ for total molecular energies and 2-3$\times$ for reaction energies.

\end{abstract}

\maketitle

\section{Introduction}

The calculation of accurate electronic energies for atoms, molecules, and extended systems is a cornerstone activity in chemical physics and physical chemistry. A quantitative description of chemical processes such as bond breaking and formation, interaction with light due to transitions to excited states, and interaction of electron/nuclear spin and external magnetic fields all require a quantum treatment of the electronic degrees of freedom and consequently, complex and time-consuming calculations of the electronic structure and energy. Among the most accurate electronic structure techniques are wavefunction-based methods, such as perturbation theory (M{\o}ller-Plesset or many-body), configuration interaction (CI), and coupled cluster (CC), among many related techniques. These methods all have the common feature of steep computational scaling with system size: apart from second-order M{\o}ller-Plesset perturbation theory (MP2) and similar approximations to CI or CC, such methods scale as at least $\mathcal{O}(N^6)$ where $N$ is a measure of system size\cite{Shavitt_Bartlett_2009,MEST_halgaker,Helgaker}.

A number of techniques to reduce the scaling of such methods through controlled approximations have been proposed over the past several decades, primarily focusing on either a localization of the molecular orbital space and exploitation of the resulting sparse structure of the Hamiltonian and wavefunction, for example the domain-localized pair natural orbital family methods,\cite{riplinger_efficient_2013,guo_communication_2018,pinski_communication_2018} and/or via a divide-and-conquer technique utilizing decomposition of the system either in real space (e.g. FMO and XSAPT)\cite{Fedorov_2018,XSAPT,hummel_low_2017} or in Hilbert space (LNO-CIM, MBBE, DEC).\cite{LNO_CIM1,LNO_CIM2,PhysRevB_106_L140204,baudin_efficient_2016} Alternatively, several methods have been proposed to reduce scaling via tensor factorization, whereby a global ``low-rank" decomposition of the Hamiltonian and/or wavefunction results in lowered computational scaling\cite{hohenstein_communication_2012,parrish_tensor_2012,benedikt_tensor_2011,hoy_comparison_2013,hummel_low_2017,schwerdtfeger_low-rank_2012}. Of course, each of these techniques necessarily introduces errors into the electronic energy and other properties. In practice, the magnitudes of the reduction in compute time and the energy error must be carefully balanced in order to obtain meaningful results on larger systems of interest.

In this work, we focus on the least squares tensor hypercontraction (LS-THC) factorization, introduced by Hohenstein et al.\cite{hohenstein_communication_2012,parrish_tensor_2012,hohenstein_tensor_2012,shenvi_tensor_2014,song_diagrammatic_2021,schutski_tensor-structured_2017,matthews_improved_2020,lee_systematically_2020,matthews_critical_2021}, for which a number of specific methods have been developed. Song and Martinez\cite{Martinez_SOS_MP2} combined scaled opposite-spin second-order Møller--Plesset perturbation theory (SOS-MP2) with THC and graphical processing units (GPUs) for efficient quantum chemistry on large molecules like proteins. These methods reduced computational scaling from quartic (fourth degree) to near-cubic (third degree) or even linear for large systems, enabling accurate, fast calculations for complex biological systems. 
Matthews\cite{matthews_critical_2021} discusses the least-squares tensor hypercontraction method for third-order Møller--Plesset perturbation theory (MP3), demonstrating promising accuracy and efficiency compared to standard density fitting. Lee et al.\cite{lee_systematically_2020} also discuss LS-THC-MP3 in the context of grids automatically generated using a weighted centroidal Voronoi partitioning. More recently, Hohenstein et al.\cite{hohenstein_rank-reduced_2022} report a full non-linear THC method (i.e. where collocation matrices are determined via direct non-linear optimization) for coupled cluster with singles and doubles (CCSD) with excellent accuracy and efficiency via multi-GPU acceleration. This latter work demonstrates that the larger errors encountered in LS-THC-MP3\cite{matthews_critical_2021,lee_systematically_2020} are not a fundamental feature of THC but an artifact of the least squares scheme, suggesting potential improvement via external correction.

Machine learning is one such powerful technique which can be used to reduce errors in computational chemistry methods. McGibbon et al. (2017)\cite{mcgibbon_improving_2017} introduced spin-network-scaled MP2 (SNS-MP2) for dimer interaction energies, achieving accuracy comparable to benchmark methods using neural network. Notably, Behler (2011)\cite{behler_neural_2011} employed feed-forward neural networks (NNs) to analyze the potential energy solution, highlighting the importance of careful training for effective results. In another paper\cite{behler_atom-centered_2011}, Behler present a high-dimensional NN based approach to evaluate potential energy surfaces by converting Cartesian coordinates onto a set of symmetry functions. Subsequent studies, including Schütt (2017)\cite{schutt_quantum-chemical_2017} and Gilmer et al. (2017)\cite{gilmer_neural_2017}, explored deep tensor neural networks and message passing neural networks, respectively, to model molecular properties and enhance computational efficiency, however these were studied for smaller molecules only.

Here, we leverage regression techniques in order to machine learn the errors which result from the least-squares tensor hypercontraction approximation to third-order M{\o}ller-Plesset perturbation theory (LS-THC-MP3). We focus on MP3 for several reasons: first, the MP3 energy may be conveniently divided into multiple physically-motivated components which can serve as independent features for machine learning (discussed further in the next section). Second, MP3 captures the majority of the necessary physics from more complex models such as coupled cluster with single and double excitations (CCSD), while being computationally simpler and more easily analyzable\cite{parrish_rank_2019,hohenstein_tensor_2012,parrish_tensor_2012,matthews_critical_2021}. Third, we have already studied the theoretical properties of LS-THC-MP3 in depth\cite{matthews_critical_2021}, in particular the relatively more inaccurate approximation of the first- or second-order wavefunction via LS-THC. The larger magnitude of these errors compared to methods which only require a decomposition of the two electron integrals\cite{matthews_critical_2021,Bertels_2019} makes LS-THC-MP3 a prime target for machine learning correction and a gateway to more general corrections of LS-THC energies. 

The main contributions of this work are: 

\begin{itemize}
    \item We apply LS-THC approximation methods to MP3 using the extensive MGCDB84 database\cite{mardirossian_thirty_2017} as a training set.
    \item We correct the error due to LS-THC in MP3 using linear and non-linear regression techniques.
    \item We compare the performance of multiple linear regression (MLR) to non-linear kernel ridge regression (KRR) models across a range of application conditions and model scenarios.
    \item We analyze the role of absolute and relative error correction in regression fitting of molecular and reaction energies.
\end{itemize}

\section{Theory}

We start by reviewing the basic theory of third-order M\o ller--Plesset theory and least squares tensor hypercontraction, then explain how the errors are induced due to approximations in LS-THC over MP3. We will then introduce our concept and approach towards application of machine learning correction of LS-THC-MP3.

\subsection{Third-Order M\o ller--Plesset Theory}

\begin{figure}
    \centering
    \includegraphics[width=8cm]{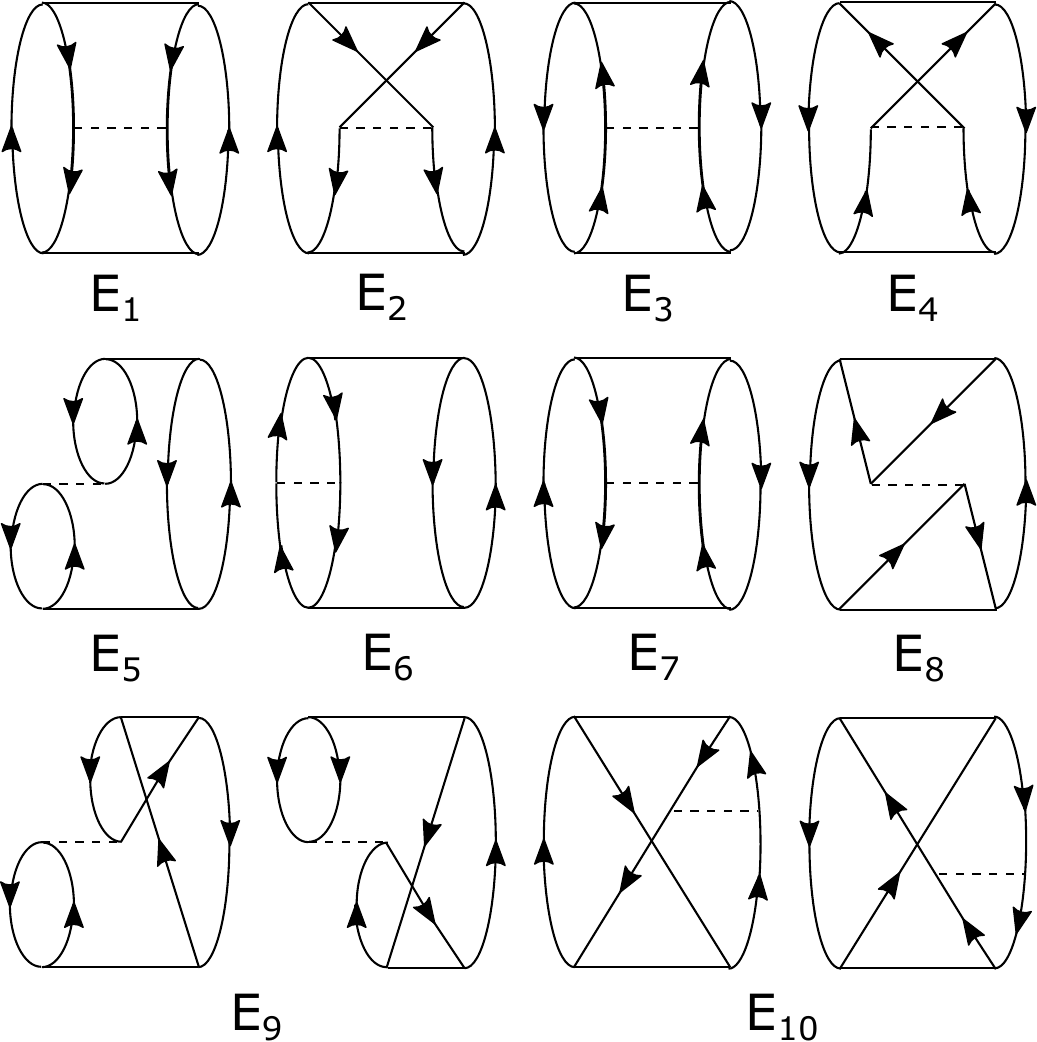}
    \caption{Goldstone diagrammatic depiction of the 10 components of the MP3 energy. Solid horizontal lines denote first-order amplitudes while dashed lines denote two-electron integrals. Note that $E_9$ and $E_{10}$ are each the sum of two Hermitian conjugate diagrams.}
    \label{fig:mp3}
\end{figure}

The total MP3 electronic energy for closed-shell systems, which includes the reference self-consistent field (SCF) and MP2 energies is given by\cite{Shavitt_Bartlett_2009},
\begin{align}
E &= E_{SCF} + E_{MP2} + E_{MP3} \nonumber \\
&= E_{SCF}
+ 2 \sum_{abij} {t^{[1]}}^{ab}_{ij} g^{ab}_{ij}
- \sum_{abij} {t^{[1]}}^{ab}_{ji} g^{ab}_{ij} \nonumber \\
&+ 2 \sum_{abijef} {t^{[1]}}^{ab}_{ij} g^{ab}_{ef} {t^{[1]}}^{ef}_{ij}
- \sum_{abijef} {t^{[1]}}^{ab}_{ji} g^{ab}_{ef} {t^{[1]}}^{ef}_{ij} \nonumber \\
&+ 2 \sum_{abijmn} {t^{[1]}}^{ab}_{ij} g^{mn}_{ij} {t^{[1]}}^{ab}_{mn}
- \sum_{abijmn} {t^{[1]}}^{ab}_{ji} g^{mn}_{ij} {t^{[1]}}^{ab}_{mn} \nonumber \\
&+ 8 \sum_{abijem} {t^{[1]}}^{ab}_{ij} g^{mb}_{ej} {t^{[1]}}^{ae}_{im}
- 4 \sum_{abijem} {t^{[1]}}^{ab}_{ij} g^{mb}_{je} {t^{[1]}}^{ae}_{im} \nonumber \\
&- 4 \sum_{abijem} {t^{[1]}}^{ab}_{ji} g^{mb}_{je} {t^{[1]}}^{ae}_{mi}
+ 2 \sum_{abijem} {t^{[1]}}^{ab}_{ji} g^{mb}_{ej} {t^{[1]}}^{ae}_{mi} \nonumber \\
&- 8 \sum_{abijem} {t^{[1]}}^{ab}_{ij} g^{mb}_{ej} {t^{[1]}}^{ae}_{mi}
+ 4 \sum_{abijem} {t^{[1]}}^{ab}_{ij} g^{mb}_{je} {t^{[1]}}^{ae}_{mi} \nonumber \\
&= E_{SCF} + E_C + E_X + \sum_{n=1}^{10} E_n \label{eq:mp3}
\end{align}
Here, $g^{pr}_{qs}$ and ${t^{[1]}}^{ab}_{ij}$ are elements of the two-electron integral tensor (the two-particle part of the normal-ordered Hamiltonian) and the first-order wavefunction amplitudes, respectively,
\begin{align}
\hat{H}_N &= \sum_{pq} f^p_q \hat{E}_{pq} + \frac{1}{2} \sum_{pqrs} g^{pq}_{rs} (\hat{E}_{pr} \hat{E}_{qs}-\delta_{rq}\hat{E}_{ps}) \\
\hat{T}^{[1]} &= \sum_{ai} {t^{[1]}}^a_i \hat{E}_{ai} + \frac{1}{2} \sum_{abij} {t^{[1]}}^{ab}_{ij} \hat{E}_{ai} \hat{E}_{bj} \\
\hat{E}_{pq} &= \hat{a}^\dagger_{p\alpha} \hat{a}_{q\alpha} + \hat{a}^\dagger_{p\beta} \hat{a}_{q\beta}
\end{align}
Specifically, the closed-shell formulation uses only the ``opposite-spin" two-electron integrals and amplitudes. For a canonical reference determinant, the first-order amplitudes are,
\begin{align}
{t^{[1]}}^a_i &= 0 \\
{t^{[1]}}^{ab}_{ij} &= \frac{g^{ab}_{ij}}{\epsilon_i + \epsilon_j - \epsilon_a - \epsilon_b} \label{eq:t2}
\end{align}
where $\epsilon_p$ are the orbital energies. The components $E_C$ and $E_X$ of the MP2 correlation energy are related to the opposite-spin and same-spin contributions used in spin-component-scaled MP2 (SCS-MP2)\cite{B709669K} by $E_{OS}=\tfrac{1}{2}E_C$ and $E_{SS}=\tfrac{1}{2}E_C + E_X$. In the MP3 correlation energy, the 10 energy components $E_n$ play a similar role. We can identify these components with distinct Goldstone diagrams\cite{matthews_critical_2021} as in Fig.~\ref{fig:mp3}.

\subsection{Least-Squares Tensor Hypercontraction}

The LS-THC approximation of MP2 and MP3 proceeds by the factorization of $g^{pq}_{rs}$ and ${t^{[1]}}^{ab}_{ij}$ into (hyper)products of five matrices\cite{hohenstein_communication_2012,parrish_tensor_2012},
\begin{align}
g^{pr}_{qs} &\approx \sum_{PQ} X_p^P X_q^P V_{PQ} X_r^Q X_s^Q \\
{t^{[1]}}^{ab}_{ij} &\approx \sum_{PQ} X_a^P X_i^P T^{[1]}_{PQ} X_b^Q X_j^Q
\end{align}
The collocation matrix $X_p^P$ is obtained by evaluating molecular orbitals $\psi_p$ at predetermined grid points $x_P$, and is shared between both the integral and amplitude factorizations. These grid points are either predefined in analogy to molecular basis sets,\cite{kokkila_schumacher_tensor_2015} or determined by pruning a larger parent grid.\cite{matthews_improved_2020,lee_systematically_2020} Only the least squares variant of THC is explored in this work, and so ``THC'' is taken to mean ``LS-THC'' in the remainder of the manuscript. In our implementation, distinct sets of grid points are used for pairs of molecular orbitals with different occupancy, i.e. virtual-virtual, virtual-occupied, and occupied-occupied. A least-squares fit is used to determined the orbital pair-specific core matrices $V^{(pq)(rs)}_{PQ}$ from density-fitted (DF) integrals,\cite{matthews_improved_2020}
\begin{align}
V_{PQ}^{(pq)(rs)}&=\sum_{P'Q'}(S^{-1}_{(pq)})_{PP'}(S^{-1}_{(rs)})_{QQ'}\sum_{pqrs}X_p^{P'}X_q^{P'}X_r^{Q'}X_s^{Q'}\tilde{g}^{pr}_{qs} \nonumber \\
&=\sum_J\left( \sum_{P'}(S^{-1}_{(pq)})_{PP'} \sum_{pq}X_p^{P'}X_q^{P'}B^{p}_{qJ}\right) \left( \sum_{Q'}(S^{-1}_{(rs)})_{QQ'}\sum_{rs}X_r^{Q'}X_s^{Q'}B^{r}_{sJ} \right) \\
\tilde{g}^{pr}_{qs} &= \sum_J B^p_{qJ}B^r_{sJ} \\
(S_{(pq)})_{PQ}&=\sum_{pq}X_p^P X_q^P X_p^Q X_q^Q
\end{align}
and similarly for core matrix $T^{[1]}_{PQ}$ through matrix-free fitting of the denominator-weighted THC-approximated integrals.\cite{matthews_critical_2021} Because the amplitudes and integrals are no longer related exactly via \eqref{eq:t2} after THC approximation, the form of \eqref{eq:mp3} as written suggests the ``THC-MP2b" and ``THC-MP3b" variants, which we further shorten to simply ``MP2b'' and ``MP3b'' as these automatically imply THC. \cite{matthews_critical_2021} In practice, we compute both the MP2a (approximation of only $g^{pr}_{qs}$) and MP2b (approximation of both integrals and amplitudes) energies, and both MP3b and MP3d energies, where the latter involves an LS-THC fit of the second-order wavefunction amplitudes. We will denote these specific energies as $E_{n,b}$, $E_{n,d}$, etc.

In order to apply machine learning techniques, we introduce a modified MP3 energy,
\begin{align}
E_{MP3}(\mathbf{c}) = \sum_{n=1}^{10} c_n E_n
\end{align}
where all $c_n = 1$ for canonical MP3. Likewise, the THC-approximated energy may be expanded in terms of individually approximated energy components, e.g.
\begin{align}
E_{MP3b}(\mathbf{c}) = \sum_{n=1}^{10} c_n E_{n,b}
\end{align}
which is potentially close to $E_{MP3}$ for some choice of $\mathbf{c}$. This construction is quite similar to SCS-MP2\cite{B709669K}, in that each energy component is assigned an independent coefficient. The use of such a modified energy formula is critical for use in non-linear machine learning in that the coefficients $c_n$ are pseudo-bounded (typically between 0 and 1, and specific constraints can be enforced), while the MP3 energy itself is scale-dependent and potentially of very different magnitude even in similarly-sized systems. Additionally, in the context of the THC approximation (which necessarily introduces some error in the MP3 energy), we can express the parameterized THC-MP3 energy as an energy difference,
\begin{align}
E_{MP3b}(\mathbf{c})- E_{MP3b} &= \sum_{n=1}^{10} c'_n E_{n,b} \nonumber \\
&= E_{\Delta MP3b}(\mathbf{c'}) \label{eq:delta-same}
\end{align}
where $c'_n = c_n-1$.

\subsection{Machine Learning}

We applied two regression techniques to determine optimal coefficients $c_n$ in either a global or a system-dependent way: multiple linear regression (MLR) and kernel ridge regression (KRR).

\subsubsection{Linear and Non-Linear Regression}

In the regression approach, the free parameters $\mathbf{C}$ of a model function $\phi(\mathbf{x};\mathbf{C})$ are determined in order to minimize an objective (or loss) function, typically the mean squared variation between the model function and the fitting data (labels) $\mathbf{Y} = \{ {\mathbf{y_1}}, {\mathbf{y_2}} ... {\mathbf{y_N} \}}$
\begin{align}
\mathbf{C} &= \underset{\mathbf{C}\in\mathbb{R}^{M\times K'}}{\arg\min} \frac{1}{N}\sum_{i=1}^N \Vert \phi(\mathbf{x}_i;\mathbf{C}) - \mathbf{y}_i\Vert^2 \nonumber \\
&= \underset{\mathbf{C}\in\mathbb{R}^{M\times K'}}{\arg\min} \frac{1}{N}\sum_{i=1}^N \Vert\hat{\mathbf{y}}_i-\mathbf{y}_i\Vert^2
\end{align}
for $N$ observable sets of $K$ features, $M$ prediction features, and $K'$ fitting parameters per prediction feature. The simplest form is multiple linear regression,\cite{chatterjee_handbook_2013} which takes the functional form,
\begin{align}
\hat{y}_{ij} = c_{j0} + \sum_{k=1}^K c_{jk} x_{ik}
\end{align}
For example, the SCS-MP2 method\cite{B709669K} may be considered a multiple linear regression of $K=K'=2$ features with zero $y$-intercept.

One version of non-linear regression uses kernels to calculate vector similarity in a high dimensional space. For example the radial basis function (RBF) or Gaussian kernel represents an exponential transform of the dot product $\langle \chi_i$, $\chi_k \rangle$ for distributions $\chi_i$ and $\chi_k$ in an infinite dimensional space,
\begin{align}
\hat{y}_{ij} = \sum_{k=1}^N c_{jk} e^{-\gamma \Vert \mathbf{x}_i - \mathbf{x}_k\Vert^2 }
\end{align}
This means we can consider such a model to have the same number of parameters as observable kernel dimensions, even though $K'=N$. Thus, the model is considered to have a continuous (infinite-dimensional) fitting space parameterized by the shape of the kernel $\mathcal{K}(\mathbf{x}_i,\mathbf{x}_j)$. The dimensionality of the RBF model may lead to over-fitting, as well as to an ill-conditioned solution, but provides a powerful capability to model non-linear relationships. The model can also be amended with ridge (also called $L_2$ or Tikhonov) regularization by adding a penalty $\alpha \sum_{i}^N \sum_{j}^M c_{ji} \hat{y}_{ij}$ to the objective function. The combination of these approaches leads to the well-known kernel ridge regression (KRR) technique.\cite{chatterjee_handbook_2013} The parameters $\gamma$ and $\alpha$ are not easily determined via closed-form solution as are the parameters $\mathbf{C}$; instead, these are treated as hyper-parameters and estimated via an additional optimization procedure such as grid search and Nelder-Mead downhill simplex optimization, vide infra.

There are other non-linear, kernel-based regression techniques such as support vector regression (SVR) and Gaussian process regression (GPR), but they differ in their underlying principles and approaches. We focus on KRR in this work as it is highly efficient and typically as performant as SVR and GPR\cite{9910959,Rasmussen2006Gaussian,PILARIO2024101056,Guo2019ShorttermPF}.  

\subsubsection{Feature Scaling}

All models discussed above benefit from transformation of the observable features $\mathbf{X}$ and labels $\mathbf{y}$ into a well-defined distribution, typically unit normal.\cite{yin2003methods, ahsan2021effect} This process of feature scaling serves to equalize the influence of different features on the objective function and to better reflect the assumptions of the KRR model. While training is done in the scaled feature (and label) space, all reported fit values are in the unscaled space, e.g. with the original units of energy. Importantly, any new data (beyond the original training set) must be scaled to the same distribution before applying the regression model, making the scaling parameters also part of the model in practice.

\section{Computational Details}

\subsection{Data Set}

As a training/test set, we use a subset of the Main Group Chemistry Database (MGCDB84)\cite{mardirossian_thirty_2017} consisting of all closed-shell systems composed of elements hydrogen through fluorine. \textcolor{review}{We focus only on this subset of MGCDB84 due to the lack of extant THC benchmarks on open-shell and third period elements, and in particular studies of necessary grid size and quality for heavier elements. We recently developed an extension of THC-MP3 to open-shell systems,\cite{doi:10.1021/acs.jctc.3c00392} but have only performed limited benchmarks thus far.} In total, this data set contains 4370 species and 2680 reactions. For each species, we computed canonical Hartree--Fock (HF), DF-MP2 and -MP3 and approximate MP2a, MP2b, MP3b, and MP3d energies (with all energies broken down into distinct diagrammatic contributions). THC calculations were performed starting with an SG-1 parent grid\cite{gill_standard_1993}, pruned via pivoted Cholesky factorization according to a tolerance parameter $10^{-\delta}$.\cite{matthews_improved_2020} Smaller $\delta$ values signify more inaccuracies and hence larger THC errors and vice versa, whereas smaller $\delta$ values also indicate a lower computational cost of the calculation. The cc-pVDZ orbital and cc-pVDZ-RI auxiliary basis sets as well as the frozen core approximation were used throughout\cite{weigend_efficient_2002,dunning_gaussian_1989,matthews_improved_2020,aquilante_unbiased_2007,rolik_efficient_2013}. We computed THC data for $\delta={1}$, ${1.25}$, ${1.5}$, ${1.75}$, and ${2}$. All calculations used a development version of the CFOUR program package.\cite{matthews_coupled-cluster_2020}

We also computed additional system-specific molecular features that affect the canonical MP3 energy and THC-MP3 energy calculations: HOMO-LUMO gap and HUMO-LOMO separation (total orbital eigenvalue span); THC goodness-of-fit measures $f_{pq}=1-\Vert B^p_{qJ} - B^q_{qJ,THC}\Vert_F/\Vert B^p_{qJ}\Vert_F$ for $pq=ab$, $ai$, and $ij$; and norms $\Vert g^{ab}_{ij}\Vert_X$ and $\Vert {t^{[1]}}^{ab}_{ij} \Vert_X$ for $X=F,\infty$ (note that the $\infty$-norm is computed as a vector norm, i.e. the maximum absolute element). This brings the total input features to 34 (ten components each of two linearly independent variants of THC: MP3b and MP3d, two components each of two variants of MP2: MP2a and MP2b, nine system-specific molecular features and one Hartree Fock energy). These additional features capture variation in the magnitude and type of interactions present in the Hamiltonian, as well as some measure of how well THC approximates the Hamiltonian.

Energy values and $g/t^{[1]}$ Frobenius norms were also normalized by dividing by the number of valence electrons to account for molecules of varying size. As the Frobenius norm is invariant to basis set rotation, we can see that by rotation to a localized basis the norm of the two-electron integrals must scale linearly with molecular size in the asymptotic limit, and a similar argument can be made for the first-order amplitudes. Deviation from asymptotic behavior for small molecules provides a non-linear feature dimension which should aid in fitting. Finally, we rescale the THC fit parameters as $\tilde{f}_{pq}=\log(1-f_{pq})$ and apply standard unit normal scaling to all feature dimensions (both training and inference) as well as the reference (canonical MP3) values for training only.

From this data we form four distinct label sets for training and/or evaluation. The {\sc Molecule} set simply contains the canonical MP3 energies for each of the 4370 species in the data set. The $\Delta${\sc Molecule} set instead tabulates the THC error directly as $\Delta E_{MP3}=E_{MP3}-E_{MP3b}$. For the {\sc Reaction} set, we combine MP3 reference energies according to the 2680 tabulated reaction schemes in our subset of MGCDB84. Here, we do not normalize the individual energies by the number of valence electrons, but rather divide the computed reaction energy $\Delta E_{rxn,MP3}$ by the total number of reactants and products, i.e., the sum of the absolute values of the stoichiometric numbers $\sum_{i=1}^{n_{species}}|\nu_i|$, for training only. Finally, the $\Delta${\sc Reaction} set combines the $\Delta${\sc Molecule} and {\sc Reaction} approaches by computing reference values which are THC reaction energy errors $\Delta\Delta E_{rxn,MP3}=\Delta E_{rxn,MP3}-\Delta E_{rxn,MP3b}$.

To summarize:
\begin{itemize}
    \item {\sc Molecule}: Directly predict MP3 energies for each molecule based on 34 features.
    \item $\Delta${\sc Molecule}: Predict $\Delta E_{MP3}$ from 34 features per molecule.
    \item {\sc Reaction}: Predict $\Delta E_{rxn,MP3}$ using the molecular energies.
    \item $\Delta${\sc Reaction}: Predict $\Delta \Delta E_{rxn,MP3}$ using the errors in molecular energies.
\end{itemize}

We performed 10-fold cross-validation by splitting the data into 10 roughly equal-sized sets or folds and training 10 independent models, holding one fold out as testing data in each case. The variation of the testing losses across folds then gives a measure of consistency within the data set and the mean of the losses gives a robust performance estimate. This serves two purposes: to verify that the model has not significantly overfit the training data (i.e., that it is generalizable) and that the distribution of the data does not contain significant outliers.

\subsection{Regression Models}

We applied both MLR ({\sc Molecule} data only) and KRR [both {\sc ($\Delta$)Molecule} and {\sc ($\Delta$)Reaction}] regression, the latter with an RBF kernel. Models for the {\sc Reaction} and $\Delta${\sc Reaction} sets were first trained on the corresponding {\sc Molecule} and $\Delta${\sc Molecule} sets and then evaluated by sequentially predicting total ($E_{MP3}$) or relative ($\Delta E_{MP3}$) energies for each species and then combining with the stoichiometric numbers to form a reaction energy prediction. Thus, the quality of the reaction energy predictions will depend on the extent of error cancellation retained or introduced in the molecular energies. The KRR hyper-parameters were tuned by initial grid search followed by Nelder-Mead downhill simplex optimization\cite{singer2009nelder}. We employed ten-fold cross-validation for both hyper-parameter search and final evaluation, where the average of the ten losses is reported.

In addition to training models with the feature set and standard scaling described above, we also trained MLR models with a reduced feature set consisting only of the 10 MP3b energy components, without feature scaling, and with a zero $y$-intercept. This model mirrors the standard SCS-MP2\cite{B709669K} in that the original energy is simply modified component-wise in order to improve accuracy, and will be known as {\sc SCS-Molecule}. Although, in this work we only target regression of the THC energy correction and not experimental or high-level computational benchmarks. Note that the {\sc SCS-$\Delta$Molecule} and {\sc MLR-$\Delta$Molecule} models are equivalent to the corresponding {\sc Molecule} models via \eqref{eq:delta-same}.

\section{Results and Discussion}

In this section, we will discuss the results of linear and non-linear regression techniques. Errors for {\sc Molecule} and {\sc $\Delta$Molecule} data are reported in \si{kcal.mol^{-1}.e^{-1}}, and in \si{kcal.mol^{-1}} for {\sc Reaction} and {\sc $\Delta$Reaction}. As mentioned earlier, lower $\delta$ values ($\sim 1$) produce intrinsically less accurate THC factorizations compared to larger values of $\delta$ ($\sim 2$), which also affects the overall accuracy and accuracy improvement of the machine learned models.

\subsection{Evaluation Criteria}
Statistical errors over the test set are reported as root mean squared errors (RMSE) for all data sets, and additionally as mean absolute errors (MAE) and mean absolute percent errors (MAPE) for {\sc Reaction} and $\Delta${\sc Reaction},
\begin{equation}
    \text{MAPE} = \frac{100\%}{N} \sum_{t=1}^{N} \left| \frac{y_i - \hat{y}_i}{y_i} \right|
\end{equation}
In order to compare machine learned models to the uncorrected THC-MP3b approximation, we also compute fractional improvement scores as,
\begin{equation}
    \text{\%IMP} = \frac{X_{MP3b} - X_{Pred}}{X_{MP3b}} * 100\%
\end{equation}
for $X=\text{RMSE}$, MAE, or MAPE. As the test set contains a wide variety of reactions/interactions, ranging from low-energy conformational rearrangements to high-energy bond dissociation and cluster binding energies, consideration of multiple error measures is critical to a full understanding of model performance.

\subsection{{\sc Molecule} and {\sc $\Delta$Molecule}}

\begin{table*}[hbt!]
\caption{Comparison of root mean squared error (RMSE) and improvement (\%IMP) results for SCS, MLR and KRR models on the {\sc Molecule} and {\sc $\Delta$Molecule} data sets (\si{kcal.mol^{-1}.e^{-1}}). Boldface numbers indicate the lowest RMSE for each value of $\delta$.}
\begin{center}
\begin{tabular}{?c?c?c|c?c|c?}
\specialrule{.1em}{.05em}{.05em} 
\multirow{2}{*}{\textbf{$\delta$ }}  &   \multirow{2}{*}{\textbf{Model}}  & \multicolumn{2}{c|}{\textbf{\sc Molecule}}  & \multicolumn{2}{c|}{\textbf{\sc $\Delta$Molecule}}   \\ 
\cline{3-6}
& & \multicolumn{1}{l|}{\centering\bf \%IMP} & \multicolumn{1}{l|}{\centering\textbf{RMSE}} & \multicolumn{1}{l|}{\centering\bf \%IMP}  &  \multicolumn{1}{l|}{\centering\textbf{RMSE}}     \\
\specialrule{.1em}{.05em}{.05em} 
\multirow{4}{*}{1} & MP3b & --- & \num{0.1337} & --- &  \num{0.1337}  \\ \cline{2-6}
 & SCS & 61\% & \num{0.0517} & 61\% & \num{0.0517}  \\ \cline{2-6}
 & MLR & 78\% & \num{0.0292} & 78\% & \num{0.0292}   \\ 
\cline{2-6}
 & KRR & 85\% & \num{0.0206} & \textbf{89\%} & \textbf{0.0153}   \\ 
 \specialrule{.1em}{.05em}{.05em} 
\multirow{4}{*}{1.25}  & MP3b & --- & \num{0.0956} & --- &  \num{0.0956}  \\ \cline{2-6}
  & SCS & 66\% & \num{0.0323} & 66\% & \num{0.0323} \\ \cline{2-6}
  & MLR & 83\%  & \num{0.0159} & 83\%  & \num{0.0159}  \\ \cline{2-6}
 & KRR  & \textbf{87\%}  & \textbf{0.0123}  & 87\%  & \num{0.0128}   \\ 
 \specialrule{.1em}{.05em}{.05em} 
\multirow{4}{*}{1.5}   & MP3b & --- & \num{0.0573} & --- &  \num{0.0573}  \\ \cline{2-6}
 & SCS & 68\% & \num{0.0183} & 68\% & \num{0.0183}  \\ \cline{2-6}
 & MLR & 84\%  & \num{0.0091} & 84\%  & \num{0.0091}   \\ \cline{2-6}
 & KRR   & \textbf{85\%}  & \textbf{0.0084} & 84\%  & \num{0.0092}  \\ 
 \specialrule{.1em}{.05em}{.05em} 
\multirow{4}{*}{1.75}   & MP3b & --- & \num{0.0316} & --- &  \num{0.0316}  \\ \cline{2-6}
 & SCS & 66\% & \num{0.0106} & 66\% & \num{0.0106}  \\ \cline{2-6}
 & MLR & 83\%  & \num{0.0053} & 83\%  & \num{0.0053}   \\ \cline{2-6}
 & KRR   & \textbf{85\%}  & \textbf{0.0048} & 84\%  & \num{0.0051}   \\ 
 \specialrule{.1em}{.05em}{.05em} 
\multirow{4}{*}{2}   & MP3b & --- & \num{0.0190} & --- &  \num{0.0190}  \\ \cline{2-6}
 & SCS & 58\% & \num{0.0079} &  58\% & \num{0.0079} \\ \cline{2-6}
 & MLR & 83\% & \num{0.0032} &   83\%  & \num{0.0032}   \\ \cline{2-6}
 & KRR   &   \textbf{85\%}  & \textbf{0.0029}  & \textbf{85\%} & \textbf{0.0029}  \\ 
\specialrule{.1em}{.05em}{.05em} 
\end{tabular}
\label{tab:Mol}
\end{center}
\end{table*}

In \Cref{tab:Mol}, RMSE and fractional improvements (\%IMP) are reported for the SCS, MLR, and KRR models on the {\sc Molecule} and {\sc $\Delta$Molecule} dat sets. For $\delta=1$, we see that the original MP3b RMSE of 0.1337 \si{kcal.mol^{-1}.e^{-1}} was reduced to 0.0517 \si{kcal.mol^{-1}.e^{-1}} using {\sc SCS-Molecule} which is a 62\% improvement over MP3b. This error was further reduced to 0.0292 \si{kcal.mol^{-1}.e^{-1}} using {\sc MLR-Molecule} which is a 78\% improvement over MP3b. {\sc KRR-Molecule} reduced this error to 0.0206 \si{kcal.mol^{-1}.e^{-1}} which is a 85\% improvement and {\sc KRR-$\Delta$Molecule} reduced the error further to 0.0153 \si{kcal.mol^{-1}.e^{-1}} which is a 89\% improvement over RMSE of MP3b. For other values of $\delta$ (=1.25, 1.5, 1.75 and 2), we see similar trend for {\sc SCS-Molecule}, {\sc MLR-Molecule}, and {\sc KRR-Molecule}, while {\sc KRR-$\Delta$Molecule} becomes slightly less effective for larger $\delta$ values. 

Considering the different machine learning models employed, the {\sc SCS-Molecule} model reduces the overall MP3b errors by approximately 60--70\%, with consistent improvement even for larger $\delta$ values. While medium values of $\delta$ achieve a larger improvement in these tests this is likely not a significant phenomenon. As with SCS-MP2, a ``straightforward'' multiple linear regression model can directly correct the errors in THC-MP3b with respect to the canonical MP3 ground state energies. From \Cref{tab:Mol}, we can see that employing a SCS correction with $\delta=1$ reduces errors to a similar level as for uncorrected THC-MP3b with $\delta=1.5$. Based on the performance scaling results of Ref.~\citenum{zhao-2023}, a reduction of $\delta$ from 1.5 to 1 corresponds to approximately a factor of 3 reduction in computational effort.

Next we can compare the RMSE results of {\sc MLR-Molecule}, which uses 34 scaled input features, to {\sc SCS-Molecule}, which uses only 10 unscaled THC components as input features. It is evident that the RMSE of {\sc MLR-Molecule} is almost half that of {\sc SCS-Molecule}, and almost one-sixth of MP3b errors across all values of $\delta$, signifying that {\sc MLR-Molecule} outperforms {\sc SCS-Molecule} when correcting the THC errors with respect to canonical MP3 values. Inclusion of additional physical features is thus an effective way to extend SCS-like approaches with low or no computational overhead. This approach could be used for novel applications of SCS, such as the present application to MP3 or other methods with distinct diagrammatic contributions, or as a modification of existing SCS-MP2 approaches. The fractional improvement for MLR is slightly lower at $\delta=1$ than for higher $\delta$ values (78\% vs. $\sim 83\%$). While the sample size is not large enough to show statistical significance, this might indicate a more significant non-linearity of the feature space at this approximation level, which we discuss further below in the context of KRR.

\begin{figure}[ht]
\begin{center}
\includegraphics[width=3.2in]{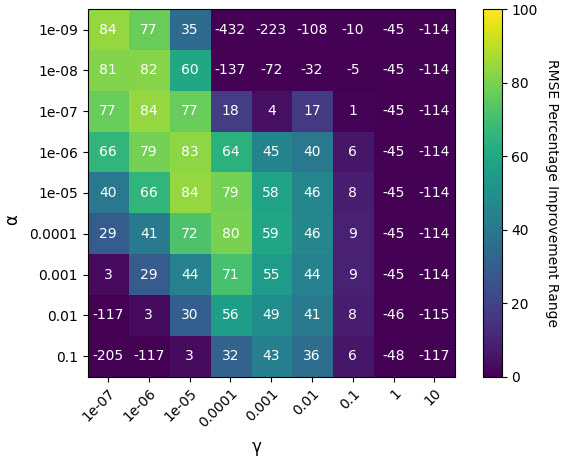} 
\includegraphics[width=3.2in]{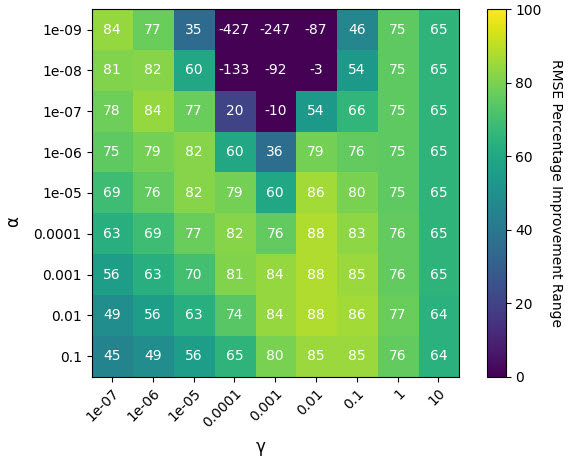}
\end{center}
\caption{RMSE \%IMP for {\sc KRR-Molecule} and {\sc KRR-$\Delta$Molecule} model with $\delta=1$ as a function of hyper-parameters $\alpha$ (regularization strength) and $\gamma$ (inverse length scale).}
\label{HM}
\end{figure}

As a first step to implement KRR, we have used grid search to identify the KRR hyper-parameters $\alpha$ and $\gamma$ for which the KRR performs best using an initial grid search followed by simplex optimization. An example of the grid search results are given in \Cref{HM}. For $\delta=1$, we can see that the maximum fractional improvement of 84\% at three [$\alpha$, $\gamma$] combinations - [1e-09, 1e-07], [1e-07, 1e-06] and [1e-05, 1e-05]. Nelder-Mead downhill simplex optimization is performed starting at point [1e-07, 1e-06] with $\alpha$ range [1e-09, 1e-05] and $\gamma$ range [1e-07, 1e-05]. The major trend in the grid search results, depicted in full in Figure S1, shows that the best fit for {\sc KRR-Molecule} is possible with very small $\alpha$ and $\gamma$ values, which is the upper left corner of the heatmap, indicating a tight, quasi-linear fit in the {\sc Molecule} dataset (the limit $\alpha,\gamma \rightarrow 0$ corresponds to the trivial prediction $\hat{y}=0$). 
However, for {\sc KRR-$\Delta$Molecule} with $\delta=1$ (\Cref{HM} and S2), a separate maximum is observed near $\gamma=0.01$ and $\alpha=0.001$, indicating significant non-linearity in the {\sc $\Delta$Molecule} dataset which can be exploited by the RBF basis. See Table S1 in the Supplemental Information for the final optimized hyper-parameters.

From \Cref{tab:Mol}, we observe that for all the $\delta$ values, the non-linear models ({\sc KRR-Molecule} and {\sc KRR-$\Delta$Molecule}) have outperformed the linear models ({\sc MLR-Molecule} and {\sc MLR-$\Delta$Molecule}), indicating the presence of at least some kernel non-linearity in feature space. Across all values of $\delta$, {\sc KRR-Molecule} and {\sc KRR-$\Delta$Molecule} achieve between 6- and 9-fold reduction in RMSE over MP3b. {\sc KRR-Molecule} and {\sc KRR-$\Delta$Molecule} perform very similarly with the exception of $\delta=1$. The higher improvement for {\sc KRR-$\Delta$Molecule} here highlights a larger degree of non-linearity in that particular dataset. As the transformation from absolute to relative errors should remove large, consistent contributions which map linearly onto input features, larger non-linearity in the error residuals is to be expected. Instead, the lower degree of non-linearity in the data for larger $\delta$ values more likely stems from an increase in incoherent noise as the errors approach zero. Compared to SCS, KRR reduces the error at $\delta=1$ to a similar level as uncorrected MP3b at $\delta=2$. The enhanced corrective power of KRR could then lead to an approximately order of magnitude reduction in required computational time for a similar level of accuracy in absolute molecular energies. \textcolor{review}{While the features used as inputs to the regression model do incur an additional computation cost, an optimized implementation of feature computation (especially the fit parameters $f_{pq}$) would render this cost negligible. In our unoptimized implementation the extra cost is less than 30\% of the total time.}

\begin{figure}
\begin{center}
\includegraphics[width=6in]{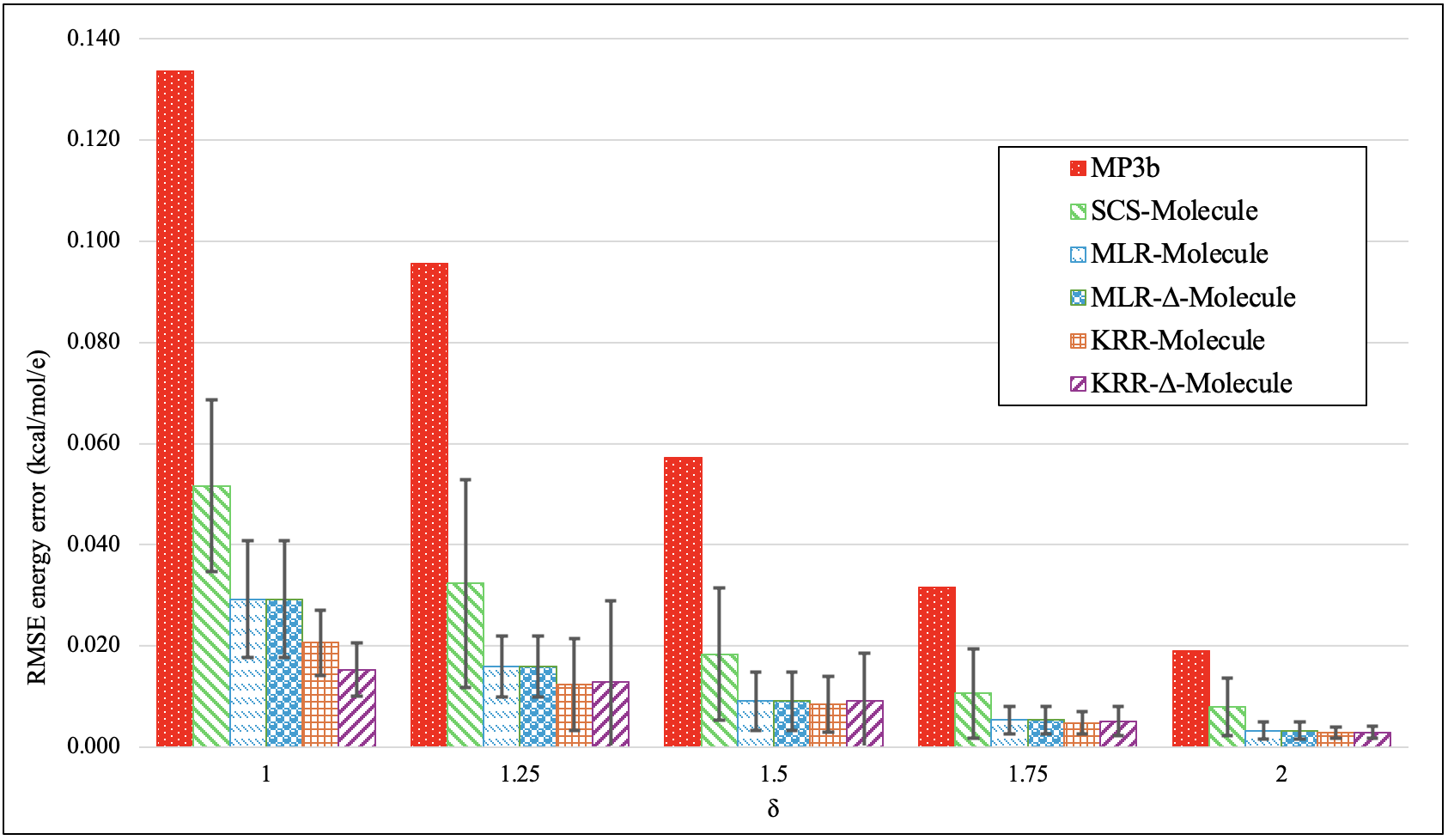} 
\end{center}
\caption{Comparative analysis of absolute energy RMSE for all models. Whiskers denote one standard deviation across the ten cross-validation sets.}
\label{fig:MolRegComp}
\end{figure}

\Cref{fig:MolRegComp} reproduces the RMSE data from \Cref{tab:Mol} in graphical form, along with ``whiskers'' representing the standard deviation across different validation sets. we can compare the RMSE of MP3b, {\sc SCS-Molecule}, {\sc MLR-Molecule}, {\sc MLR-$\Delta$Molecule}, {\sc KRR-Molecule} and {\sc KRR-$\Delta$Molecule} for different $\delta$ values. Longer whiskers indicate that the RMSE from 10 fold validation have larger variation and smaller whiskers indicate that the 10 RMSE results are tightly clustered around the mean, suggesting a more reliable and stable model performance across data points. Larger standard deviations more likely represent variability in the dataset rather than overfitting, as overfitting (especially in the KRR model) would manifest as poor test set predictions for all folds, lowering the average improvement as well.

We also observe that although {\sc SCS-Molecule} has reduced the MP3b errors by almost half, the variation of prediction errors are very large; this variation in the RMSE has been significantly reduced using higher dimensional and scaled dataset using {\sc MLR-Molecule} and {\sc MLR-$\Delta$Molecule}. The variations in {\sc KRR-Molecule} is smaller as compared to {\sc KRR-$\Delta$Molecule}, presumably due to a non-uniform distribution of the non-linearity in the {\sc $\Delta$Molecule} dataset. However, for $\delta=1.25$, we see an abrupt change in behavior in the standard deviations: the variation of the non-linear models (KRR) is higher than that of the linear model (MLR), indicating the presence of outliers or uneven fit quality. The 1-$\sigma$ ranges of different models for the {\sc Molecule} and {\sc $\Delta$Molecule} datasets overlap, indicating similar overall statistical performance. For example in case of {\sc MLR-Molecule} and {\sc MLR-$\Delta$Molecule} for all $\delta$ values, and {\sc KRR-Molecule} and {\sc KRR-$\Delta$Molecule} for $\delta=2$.
It is evident from Figure \ref{fig:MolRegComp}, that the non-linear regression models have reduced the THC-MP3b errors significantly for all $\delta$ values.

\subsection{{\sc Reaction} and {\sc $\Delta$Reaction}}

\begin{table*}[ht]
\caption{Comparison of RMSE (\si{kcal.mol^{-1}}), MAE (\si{kcal.mol^{-1}}), MAPE (\%) and their fractional improvements (\%IMP) for {\sc KRR-Reaction} and {\sc KRR-$\Delta$Reaction} models. Here, R represents {\sc KRR-Reaction} and $\Delta$R represents {\sc KRR-$\Delta$Reaction}. Boldface values denote the best model for each value of $\delta$, if evident.}
\begin{center}
\begin{tabular}{?c?c?c|c?c|c?c|c?}
\specialrule{.1em}{.05em}{.05em} 
{$\delta$ }  & \textbf{Model}  &   \textbf{\%IMP} & \textbf{RMSE} &  \textbf{\%IMP} & \textbf{MAE}  &   \textbf{\%IMP} & \textbf{MAPE} 
\\
\specialrule{.1em}{.05em}{.05em} 
\multirow{3}{*}{1} & MP3b & --- & \num{1.5711} & --- & 0.9374 & --- & 27.51\%   \\ \cline{2-8}
 & R & 38\% & \num{0.9835} & 29\% & \num{0.6625} & 34\% & {18.14\%}   \\ \cline{2-8}
& $\Delta$R  & \textbf{64\%} & \textbf{0.5732} & \textbf{59\%} & \textbf{0.3885} & \textbf{70\%} & \textbf{8.31\%}   \\ 
\specialrule{.1em}{.05em}{.05em} 
\multirow{3}{*}{1.25} & MP3b & --- & \num{0.8869} & --- & 0.4853 & --- & 11.33\%   \\ \cline{2-8}
 & R & 55\% & \num{0.4050} & 45\% & \num{0.2675} & 27\% & {8.31\%}   \\ \cline{2-8}
& $\Delta$R  & \textbf{65\%} & \textbf{0.3078} & \textbf{56\%} & \textbf{0.2153} & \textbf{53\%} & \textbf{5.36\%}   \\ 
\specialrule{.1em}{.05em}{.05em} 
\multirow{3}{*}{1.5} & MP3b & --- & \num{0.5975} & --- & 0.2998 & --- & 10.77\%    \\ \cline{2-8}
 & R & 51\% & \num{0.2960} & 37\% & \num{0.1894} & 48\% & {5.63\%}   \\ \cline{2-8}
& $\Delta$R  & 53\% & {0.2828} & {36\%} & {0.1915} & {36\%} & {6.93\%}   \\ 
\specialrule{.1em}{.05em}{.05em} 
\multirow{3}{*}{1.75} & MP3b & --- & \num{0.4036} & --- & 0.1943 & --- & 4.35\%  \\ \cline{2-8}
 & R & 51\% & \num{0.1969} & 40\% & \num{0.1174} & 13\% & {3.77\%}   \\ \cline{2-8}
& $\Delta$R  & 51\% & \num{0.1981} & 39\% & \num{0.1182} & 12\% & {3.81\%}   \\ 
\specialrule{.1em}{.05em}{.05em} 
\multirow{3}{*}{2} & MP3b & --- & \num{0.2814} & --- & 0.1327 & --- & 2.81\%    \\ \cline{2-8}
 & R & \textbf{57\%} & \textbf{0.1218} & \textbf{48\%} & \textbf{0.0691} & \textbf{44\%} & \textbf{1.56\%}   \\ \cline{2-8}
& $\Delta$R  & 53\% & \num{0.1322} & 45\% & \num{0.0736} & 41\% & {1.66\%}   \\ 
\specialrule{.1em}{.05em}{.05em} 
\end{tabular}
\label{tab:KRRReact}
\end{center}
\end{table*}

\Cref{tab:KRRReact} presents RMSE, MAE, and MAPE results for the {\sc KRR-Reaction} and {\sc KRR-$\Delta$Reaction} models. For $\delta=1$, the RMSE of MP3b is 1.5711 \si{kcal.mol^{-1}}, which was reduced by {\sc KRR-Reaction} to 0.09835 \si{kcal.mol^{-1}} which is a 38\% improvement and by {\sc KRR-$\Delta$Reaction} to 0.05732 \si{kcal.mol^{-1}} which is a 64\% improvement. 
It is clear that using {\sc KRR-Reaction} and {\sc KRR-$\Delta$Reaction}, the reduced THC errors from molecular energies were able to yield reduced THC errors in reaction energies from corrections to isolated molecular energies alone, although the extent of improvement is somewhat less.

We also see that in terms of RMSE, {\sc KRR-$\Delta$Reaction} outperforms {\sc KRR-Reaction} for $\delta=1$ and 1.25, whereas both models perform similarly for $\delta=1.5$ and 1.75, and the relationship reverses for $\delta=2$. We see that {\sc KRR-Reaction} has performance improvement ranging between 38\% for $\delta=1$ and 57\% for $\delta=2$, and {\sc KRR-$\Delta$Reaction} has performance improvement ranging between 51\% for $\delta=1.75$ and 65\% for $\delta=1.25$, with all other $\delta$ between these ranges. We also see that the performance of both the models ``peak'' at $\delta=1.25$. Also, for $\delta=1.5$, 1.75 and 2, the RMSE of both the models has marginal difference. Hence we will also look for how both the models perform on an absolute scale. The better performance of {\sc KRR-$\Delta$Reaction} for small $\delta$ values highlights that for loose THC approximations, the larger scale of the energy corrections offers more scope of improvement through non-linear fitting, along with a potential increase in non-linearity due to cancellation between molecular energies. As with the {\sc Molecule} data set, KRR provides the most relative improvement for large initial errors. However, the larger improvement of the KRR model for molecular energies compared to reaction energies points to inadequate error cancellation in the {\sc KRR-Molecule} models. Even though it uses physical features, the KRR model is not physics-based and so cannot ``anticipate'' the necessary cancellation required to most accurately predict residual reaction or interaction energies. Instead, the KRR model introduces more incoherent random errors which do not reliably cancel.

We see that for all $\delta$ values, the RMSE are consistent with MAE and MAPE, except for $\delta=1.75$. In this case, the MP3b MAPE reduces suddenly from 10.77\% to 4.35\%, while the KRR models show a smoother decrease in MAPE with increasing $\delta$. However, with the exception of this point, the MAE and MAPE fractional improvement of both the models are roughly comparable across all values of $\delta$ and obtain a 2--$3\times$ reduction in error for predicted reaction energies.

\section{Conclusions}

From the above results, we can conclude that all tested machine learning models provide significant reduction in LS-THC-MP3b errors at little to no additional computational cost. Multiple linear regression based on a spin-component-scaling approach is effective, yielding an improvement factor of 58--68\% over MP3b, but introducing additional physical input features further improved to 78--84\%. The significant effect of introducing additional energy-based and non-energy-based input features suggests that ``traditional'' SCS approaches could also be improved, and that other methods admitting a diagrammatic decomposition could be corrected using an SCS-like scheme.

However, introducing a non-linear regression model with KRR reduced the THC errors further, indicating non-linearity present in the {\sc Molecule} and {\sc $\Delta$Molecule} data sets, that was exploitable using a radial basis function basis. The performance of {\sc KRR-Molecule} is between 85\% to 87\%, and for {\sc KRR-$\Delta$Molecule} is in the range 84--89\%. The performance of both the models varied slightly depending on $\delta$ values, with {\sc KRR-$\Delta$Molecule} achieving the best performance at $\delta=1$, which corresponds to the loosest THC approximation. These results highlight the ability of KRR to more accurately correct large, non-linear errors in comparison to MLR.

We also saw that by training regression models to correct THC errors in molecules (the {\sc Molecule} data set), we can also reduce THC errors in derived reaction and interaction energies (the {\sc Reaction} data set) utilizing the concept of error cancellation between reactants and products in a reaction. The performance of {\sc KRR-Reaction} ranges from 38\% to 57\% improvement over MP3b and {\sc KRR-$\Delta$Reaction} ranges from 51\% to 65\% improvement over MP3b. The performance of both the models varied depending on $\delta$ values. The lower improvement of reaction energies compared to molecular energies highlights the random distribution of errors produced by KRR which cannot exploit physical similarity of reactants and products.

\textcolor{review}{While this study focused only on the closed-shell, second period subset of MGCDB84, and models trained on this subset are not likely to generalize to the wider data set, our observations support the conclusion that retraining of regression models on more diverse systems is likely to also result in substantial reduction of THC errors. Importantly, the relative quality of grids for light and heavy elements must be balanced in order for the grid size parameter $\delta$ to be meaningful.}

In summary, we show that simple regression models are highly effective in correct errors inherent in the LS-THC approximation, with non-linear kernel ridge regression providing the most improvement, more than 84\% for total molecular energies, but for the {\sc Reaction} data set, the improvement was less than 65\%. This situation highlights both the promise of regression for overcoming the accuracy limitations of LS-THC, but also the limitations of the regression method itself for studying subtle changes in  reaction energies which depend on error cancellation. 

\begin{acknowledgments}
This work was supported in part by the National Science Foundation (grants OAC-2003931 and CHE-2143725). Computational
resources for this research were provided by SMU’s O’Donnell Data Science and Research
Computing Institute.
\end{acknowledgments}

\section*{Supplemental Information}
A supplemental information file (.pdf) is provided including the following information:
\begin{itemize}
    \item Full KRR hyper-parameter grid search data.
    \item Optimized $\alpha$ and $\gamma$ hyper-parameters.
\end{itemize}

\bibliography{ANN-THC}

\end{document}